\documentclass{basi}
\usepackage{savesym}
\usepackage{amsmath}
\savesymbol{iint}
\usepackage{txfonts}
\restoresymbol{TXF}{iint}

\begin{document}

\title[Iron line in OAO~1657-415]{Investigation of iron emission lines in the eclipsing high mass X-ray binary pulsar OAO~1657-415}

\author[G.K. ~Jaisawal and S.~Naik]{Gaurava K. Jaisawal$^1$\thanks{email:
       \texttt{gaurava@prl.res.in}} and Sachindra Naik$^1$\thanks{email:
       \texttt{snaik@prl.res.in}}\\
        $^1$Physical Research Laboratory, Navrangapura,
           Ahmedabad 380009, Gujarat, India}

\pubyear{2014}
\volume{42}
\pagerange{\pageref{firstpage}--\pageref{lastpage}}

\date{Received 2014 August 13; accepted 2014 December 01}
\maketitle
%

\begin{abstract}
We present the results obtained from timing and spectral studies of high 
mass X-ray binary pulsar OAO~1657-415 using a {\it Suzaku} observations in 
2011 September. X-ray pulsations were detected in the light curves up to 
$\sim$70 keV. The continuum spectra during the high- and low-flux regions in 
light curves were well described by high energy cutoff power-law model along 
with a blackbody component and iron fluorescent lines at 6.4 keV and 7.06 keV. 
Time resolved spectroscopy was carried out by dividing the entire observations 
into 18 narrow segments. Presence of additional dense matter at various orbital 
phases was confirmed as the cause of low-flux regions in the observations. Presence 
of additional matter at several orbital phases of the pulsar was interpreted as 
due to the inhomogeneously distributed clumps of matter around the neutron star. 
Using clumpy wind hypothesis, the physical parameters of the clumps causing the 
high- and low-flux episodes in the pulsar light curve were estimated. The 
equivalent width of iron emission lines was found to be significantly 
large at certain orbital phases of low-flux segments. We investigated the iron 
line emitting regions and suggest the existence of neutral and ionized iron atoms 
in emission sites that are located within the accretion radius.     
\end{abstract}


\section{Introduction}
Accretion powered high mass X-ray binary pulsar OAO~1657-415 was discovered 
by Copernicus satellite (Polidan et al. 1978). Initially, the object was 
incorrectly identified with the massive binary V861~Sco and the
system interpreted as a black hole binary. However, the observations with HEAO~1 
and Einstein 
observatories were used to determine the precise position of the source that
ruled out V861~Sco as the companion of OAO~1657-415. Using the same data, 
a pulsation period of 38.22 s was detected in the pulsar (White \& Pravo 1979; 
Parmar et al. 1980). Long-term monitoring of the pulsar with the Burst and 
Transient Source Experiment (BATSE) onboard Compton Gamma Ray Observatory 
(CGRO) confirmed the system as an eclipsing binary (Chakrabarty et al. 1993). 
The orbital parameters derived from the BATSE observations are: orbital 
period P$_{orb}$ = 10.444$\pm$0.004 d, duration of the eclipse = 1.7d, $e$ 
= 0.104$\pm$0.005, $a_x$sin~$i$ = 106$\pm$0.5 lt-sec and $\omega$ = 
93$^\circ\pm5^\circ$. Using these orbital parameters, the mass and radius of 
companion were estimated to be in the range of 14--18 M$_\odot$ and 25-32 
R$_\odot$, respectively (Chakrabarty et al. 1993). However, the counterpart 
was not detected at the $Chandra$ position of the pulsar in the deep optical 
imaging of the field with a limitings magnitude of V$<$23, near infrared (NIR)
imaging revealed a highly reddened B-type supergiant star as binary companion 
of OAO~1657-415 (Chakrabarty et al. 2002). The distance of binary system was 
estimated to be 6.4$\pm$1.5 kpc. On the basis of NIR spectroscopy, the spectral 
class of the binary companion was refined as Ofpe/WNL type star (Mason et al. 
2009). Such stars are known to be the transitional objects between the main 
sequence and Wolf-Rayet stars. 

In high mass X-ray binary (HMXB) systems, mass transfer occurs from 
the binary companion to the neutron star through the capture of stellar 
wind, accretion through the Be circumstellar disk or the Roche lobe overflow.
Based on the mechanism of mass accretion, the supergiant X-ray binaries 
(SGXBs; a subclass of HMXBs) are classified into disk-fed and wind-fed SGXB.
HMXBs exhibit a definite positions in the spin period versus orbital period 
diagram (Corbet 1986) depending on the type of mass accretion. The wind-fed 
SGXBs are distributed in the form of horizontal line in Corbet diagram with 
exceptions of OAO~1657-415 and 2S~0114+65. However, the three known disk-fed SGXBs 
such as Cen~X-3, SMC~X-1 and LMX~X-4 show anti-correlation between spin and 
orbital period. The moderate values of spin and orbital period of OAO~1657-415 
make the pulsar as an unique transitional source from wind-fed to disk-fed SGXB. 

Spin period evolution of OAO~1657-415 was studied by using $BATSE$ and
{\it Rossi X-ray Timing Explorer (RXTE)} observations (Baykal 1997;
Bildsten et al. 1997; Baykal 2000). $BATSE$ observations of the pulsar 
revealed strong and stochastic variability in the spin period of the pulsar. 
Alternating steady spin-up and spin-down episodes lasting for 10-200 d, 
as seen in Cen~X-3, were detected in OAO~1657-415 (Bildsten et al. 1997).
These episodes of frequent torque reversals cannot be explained only by 
wind accretion, rather possibility of the formation of a temporary
accretion disk was considered to explain the observed changes in spin
period history (Baykal 1997). $RXTE$ observations of the pulsar in 1997 
August covering a binary orbit showed an extended spin-down episode (Baykal 
2000). During these observations, a marginal correlation was seen between
X-ray luminosity and change in the pulse frequency of the pulsar. This 
positive correlation suggested that the disc formed in the spin-down episode 
is in the prograde direction. Study of the pulse frequency history of OAO~1657-415 
by using nearly 19 years of data obtained from {\it BASTE} and {\it Gamma-Ray 
Burst Monitor (Fermi/GBM)} observations inferred the occurrence of two modes of 
accretion in the pulsar. Presence of a stable accretion disk was suggested during 
flux correlated steady spin-up episodes (disk wind accretion) whereas the other 
type is during flux uncorrelated spinning down of the pulsar at a lesser rate 
(wind accretion) (Jenke et al. 2012).

The pulsar being at low galactic latitude, the X-ray spectrum shows significant
photoelectric absorption at soft X-rays with an equivalent hydrogen column density 
of $\geq$10$^{23}$ atom cm$^{-2}$ (Parmar et al. 1980). However, $Ginga$ observations 
of the pulsar showed the presence of soft X-ray excess below $\sim$3 keV (Kamata et 
al. 1990). Apart from the soft X-ray excess, an iron fluorescence emission line at 
$\sim$6.6 keV was also detected in the $Ginga$ spectrum of the pulsar. {\it ASCA}
observation of the pulsar near mid-eclipse time revealed that though the continuum
was weak, the 6.4 keV iron fluorescent emission line was dominant compared to 6.7 keV. 
Another long {\it ASCA} observation covering non-eclipse through mid-eclipse phase of the
pulsar showed that the out-of-eclipse high intensity spectrum had an absorbed 
continuum component along with 6.4 keV and 7.06 keV iron emission lines (Audley et
al. 2006). Broad-band spectrum of the pulsar in 1.0-100 keV energy range, obtained 
from $BeppoSAX$ observation, was found to be well described by a cutoff power law or 
power-law modified by a high energy cutoff continuum model along with an iron 
fluorescence line at 6.5 keV (Orlandini et al. 1999). Also there was a hint of 
the presence of a cyclotron absorption feature at $\sim$36 keV as required by one 
of the above continuum model, the absence of such feature in the model independent 
Normalized Crab ratio - a technique used to identify cyclotron scattering resonance 
features in the spectra of X-ray pulsars - made the detection inconclusive (Orlandini 
et al. 1999). {\it INTEGRAL} observations also did not find any signature of the 
cyclotron absorption features in the broad-band (6-200 keV range) spectrum of 
OAO~1657-415 (Barnstedt et al. 2008). 

For a detailed study of the characteristics properties of iron emission lines and 
broad-band spectral properties, the eclipsing HMXB pulsar OAO~1657-415 was observed 
with $Suzaku$ on 2011 September 26. Though this observation was recently used to
describe the timing and spectral properties of the pulsar (Pradhan et al. 2014), we
investigated the properties of iron emission lines at different orbital phases of
the pulsar and tried to interpret the results by applying clumpy wind model. 
The results obtained from our study are described in the paper.

\section{Observations}
$Suzaku$, the fifth Japanese X-ray astronomy satellite, was launched by Japan 
Aerospace Exploration Agency (JAXA) on 2005 July 10 (Mitsuda et al. 2007). It 
covers very wide energy range (0.2-600 keV) with two sets of detectors, X-ray 
Imaging Spectrometer (XIS; Koyama et al. 2007) and Hard X-ray Detector 
(HXD; Takahashi et al. 2007). Three front illuminated (XIS-0, XIS-2, XIS-3) 
and one back illuminated (XIS-1) CCD detectors are located at focal point of 
their respective X-ray telescopes. Field of view (FoV) of XIS is 
17$'$.8$\times$17$'$.8 in full window mode. Effective areas of front illuminated 
and back illuminated detectors in full window mode are 340 cm$^2$ and 390 cm$^2$ 
at 1.5 keV, respectively. HXD detector comprises of two sets of instruments - 
PIN and GSO. PIN is silicon diode detector working in 10--70 keV energy range 
whereas GSO is scintillator detector that covers 40--600 keV energy range. The 
effective areas for PIN and GSO are 145 cm$^2$ at 15 keV and 315 cm$^2$ at 100 keV, 
respectively. FoV of PIN and GSO (below 100 keV) is $34'\times34'$. The 
best time resolution which can be achieved in normal mode of HXD/PIN is 61 
$\mu\mathrm{s}$ (Kokubun et al. 2007).

\begin{figure}
\centering
\includegraphics[height=3.in, width=2.in, angle=-90]{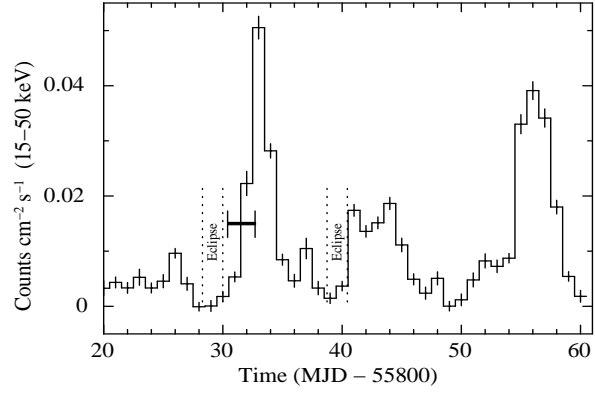}
\caption{{\it Swift}/BAT light curve of OAO~1657-415 in 15--50 keV energy range,
from 2011 September 16 (MJD 55820) to 2011 October 26 (MJD 55860). The horizontal
mark shows the duration of the $Suzaku$ observation of the pulsar. The durations
of the eclipse are marked in the figure.}
\label{sw}
\end{figure}

OAO~1657-415 was observed by {\it Suzaku} observatory on 2011 September 26--28 
for an exposure of $\sim$85 ks. The out-of-eclipse {\it Suzaku} observation 
of the pulsar covered about 20\% of the binary orbit in 0.12-0.34 orbital phase 
range (considering mid-eclipse time as phase 0). The archival data of processing 
version 2.7.16.30 were used in the present analysis. $Suzaku$ observation was 
carried out in ``XIS nominal'' pointing mode in which source is kept at the center 
of XIS detector. The effective exposures for XIS and HXD/PIN were $\sim$85 ks and 
$\sim$75 ks, respectively. XIS detectors were operated in ``1/4 window mode'' 
with a FoV of 17$'$.8$\times$4$'$.4 and time resolution of 2 s. {\it Swift}/BAT 
light curve of the pulsar from 2011 September 16 to 2011 October 26 is shown in 
Figure~\ref{sw} covering the energy range in 15--50 keV. The horizontal line in 
the figure shows the duration of the $Suzaku$ observation of the pulsar. In the 
figure, the duration of the eclipse, estimated by using the orbital parameters 
given by Jenke et al. (2012), are also marked. 
   
Heasoft package ver 6.12 was used in the present analysis. Calibration database
files released on 2012 February 10 (for XIS) and 2011 September 13 (for HXD) were used 
for reprocessing of $Suzaku$ data. Unfiltered files were reprocessed by using the task 
`aepipeline'. Barycentric correction was applied on the reprocessed clean event 
files by applying `aebarycen' task. Source light curves and spectra were extracted 
from XIS event data by considering a circular region of $4'$ radius with source at the
center, located on $8'$ long frame of the ``1/4 window mode'' strip. Background light 
curves and spectra were extracted from XIS event data by considering source-free 
circular regions of $2'$ radius centered at a distance of $6'$ away the source center 
on the ``1/4 window'' strip. The response and effective area files for each XIS were 
generated by using tasks `xisrmfgen' and `xissimarfgen'. HXD/PIN light curves and spectra 
of the pulsar were extracted from corresponding event files by using XSELECT package of 
FTOOLS. For HXD/PIN background, simulated non-X-ray background (NXB) event files 
corresponding to 2011 September 26 were used to estimate the 
NXB\footnote{http://heasarc.nasa.gov/docs/suzaku/analysis/pinbgd.html} and the 
cosmic X-ray background was simulated as suggested by the
instrument team\footnote{http://heasarc.nasa.gov/docs/suzaku/analysis/pin\_cxb.html}.
The response files released on 2011 June 01 was used for HXD/PIN in the spectral
analysis. Data from XIS-0, XIS-1, XIS-3 and HXD/PIN detector are used in our 
analysis. We have not included the GSO data due to low signal to noise ratio and 
high background in the hard X-ray energy ranges.  

\begin{figure}
\centering
\includegraphics[height=3.1in, width=4.1in, angle=-90]{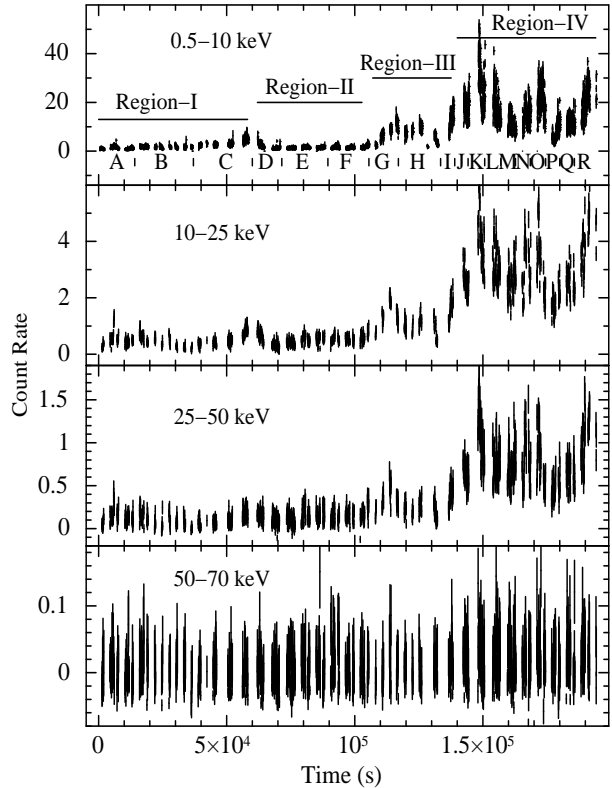}
\caption{Light curves of OAO~1657-415 in 0.5-10 keV (combined data from 
XIS-0, 1 \& 3), 10-25 keV, 25-50 keV and 50-70 keV (HXD/PIN) energy ranges 
obtained from the $Suzaku$ observation of the pulsar in 2011 September. Depending 
on the variable X-ray flux during the observation, the data were divided into four 
broad regions as marked in the figure (top panel). The acronyms in the top panel 
A to R show the segments of the data used for detailed time resolution spectroscopy 
as described in the later part of the paper. The other panels show the presence/absence 
of narrow and extended low flux levels during the $Suzaku$ observation. All the light 
curves are background subtracted.}
\label{xislc}
\end{figure}

\section{Results}
During the $Suzaku$ observation, OAO~1657-415 was found to exhibit flux 
variability on several time scales. The combined XIS light curve in 
0.5-10 keV range showed the presence of two extended low intensity 
segments and two high intensity segments as shown in Figure~\ref{xislc}. 
The extended low flux segments were found in the light curves up to 
as high as $\sim$50 keV. To investigate the properties of the pulsar 
at different flux levels, the entire observation was divided into four 
broad regions and marked as Regions-I, II, III, \& IV in the top panel of 
Figure~\ref{xislc}. Light curves were extracted from XIS and HXD/PIN data 
at 2 s and 1 s time resolutions, respectively, for all four segments. 
The light curves were corrected for the orbital motion of the neutron
star in the binary system. As the {\it Suzaku} observation covered
about 0.2 orbital phase of the binary system, the light curves were
corrected for the orbital motion by using the ephemeris given by Jenke 
et al. (2012). The barycentric and orbital motion corrected light curves 
were further used in the timing analysis of the pulsar. We have 
examined the energy resolved pulse profiles for all four regions by folding 
the corresponding background subtracted light curves on the average  
pulse period.  It was found that pulsations were present in the light 
curves up to as high as $\sim$70 keV. The shape of the pulse profiles i.e. 
complex shape up to $\sim$25 keV because of the presence of several narrow 
absorption dips, were found to be similar to that reported in Pradhan 
et al. (2014).

To study the changes in spectral properties of OAO~1657-415 in various segments
of high and low flux levels, we extracted source and background spectra from 
each of XIS and PIN detectors and generated corresponding response matrices as 
described earlier. To improve the signal-to-noise ratio, the spectral data obtained 
from XIS instruments were re-binned by a factor of 30 in 1.8-6 keV range, 12 in 6-8 keV 
range and 25 in 8-10 keV range. Similarly, HXD/PIN data were re-binned by a factor of 12 
in 10-15 keV, 6 in 15-50 keV  and 10 in 50-70 keV energy range. For a detailed 
investigation of spectral evolution of the pulsar at different flux levels during 
the $Suzaku$ observation, we carried out the simultaneous spectral fitting of data 
obtained from XIS-0, XIS-1, XIS-3 and HXD/PIN in 1.8-70 keV energy range. Before 
spectral fitting, appropriate background subtraction was done from 
the data obtained from all the detectors. Because of the presence of known features 
of Si and Au, XIS data in 1.7-1.9 keV and 2.2-2.4 keV ranges were ignored from 
spectral fitting. All the model parameters were tied during spectral fitting except 
the detector normalizations which were kept free. In the beginning, we tried to fit 
the spectral data with standard continuum models such as high energy cutoff 
power law model, Negative and Positive power law with Exponential cutoff (NPEX) 
model along with interstellar absorption and iron emission line at 6.4 keV. 
Simultaneous spectral fitting of the XIS and HXD/PIN data during the 
high flux level of Region-III with high energy cutoff power law continuum 
model along with interstellar absorption and a Gaussian function for 6.4 keV 
iron emission line yielded a merely acceptable fit with a reduced $\chi^{2}$ 
of 1.91 for 156 degrees of freedom (dof). Presence of an emission line like 
feature at $\sim$7.1 keV allowed us to add another Gaussian function to the
spectral model. Addition of the second Gaussian function improved the spectral
fitting further with a reduced $\chi^{2}$ of 1.51 (153 dof). A blackbody component 
was added to the spectral model for the soft excess providing the best fit model 
with a reduced $\chi^{2}$ of 1.19 (151 dof). The values of relative instrument
normalizations of the three XISs and PIN detectors obtained from the spectral fitting
are 1.00, 0.87, 0.95 \& 1.14 for XIS-0, XIS-1, XIS-3 \& HXD/PIN, respectively and 
are comparable with the values obtained at the time of detector calibration.
We attempted to fit the pulsar spectrum with NPEX continuum model along with 
interstellar absorption and two Gaussian functions for iron emission lines. 
NPEX model along with other components improved the spectral fitting marginally 
with a reduced $\chi^{2}$ of 1.13 (151 dof). However, the value of HXD/PIN detector
normalization was found to be very high (1.5). Therefore, we discarded the NPEX 
continuum model in our spectral fitting. While fitting the XIS and HXD/PIN spectra 
during all other regions, we found that the high energy cutoff power law continuum 
model fitted data better than NPEX continuum model. The mathematical form of the 
high energy cutoff power law model is: 

\begin{equation}
I(E)= \left\{ \begin{array}{lc}
E^{-\gamma} & (E\le E_c) \\
E^{-\gamma}\exp\left(-\frac{E-E_c}{E_f}\right) & (E>E_c)
\end{array}\right.
\end{equation}
where $E_c$ and $E_f$ are cutoff energy and  folding energy, respectively.

\begin{table}
\centering
\caption{Best-fit parameters (with 90$\%$ errors) for the four regions.}

\begin{tabular}{lllll}
\hline
Parameter      	                                           	&\multicolumn{4}{c}{Value} 	 \\
                                      &Region-I           &Region-II          &Region-III          &Region-IV\\
\hline
N$_{H}$ (10$^{22}$ atoms cm$^{-2}$)   &51.33$\pm$4.09     &75.74$\pm$5.91     &24.81$\pm$1.67      &28.45$\pm$0.93 \\
Power-law index                       &1.04$\pm$0.12      &1.12$\pm$0.07      &1.16$\pm$0.07       &0.98$\pm$0.04 \\
Norm.$^a$ of Power-law (10$^{-2}$)    &1.24$\pm$0.30      &1.41$\pm$0.43      &3.78$\pm$0.60       &5.83$\pm$0.48 \\
$kT_{BB}$ (keV)                       &0.33$\pm$0.04      &0.30$\pm$0.05      &0.32$\pm$0.03       &0.32$\pm$0.01 \\
Norm.$^b$ of $kT_{BB}$ (10$^{-2}$)    &2.04$\pm$1.39      &18.43$\pm$14.31    &1.82$\pm$0.83       &6.99$\pm$1.56  \\
High energy cut-off (keV)	            &16.74$\pm$2.88     &20.81$\pm$1.34     &19.22$\pm$2.17      &18.54$\pm$0.57 \\
Folding energy  (keV)                 &19.38$\pm$1.85     &21.38$\pm$2.61     &23.05$\pm$3.01      &21.69$\pm$0.81 \\
\\
Fe K$_\alpha$ line Energy (keV)		    &6.413$\pm$0.006	  &6.411$\pm$0.003	  &6.411$\pm$0.005	   &6.420$\pm$0.004 \\
Width of Fe K$_\alpha$ (keV)          &0.01$\pm$0.03      &0.01$\pm$0.01      &0.01$\pm$0.01       &0.03$\pm$0.02 \\
Norm.$^c$ of Fe K$_\alpha$			      &3.92$\pm$0.28	    &18.74$\pm$1.40     &9.20$\pm$0.46	     &20.94$\pm$0.55 \\
Fe K$_\alpha$ eq. width (eV)          &218.34$\pm$14.71   &1069.38$\pm$79.61  &211.46$\pm$10.49    &221.42$\pm$5.82\\
Fe K$_\alpha$ line flux$^d$           &1.61$\pm$0.09      &4.94$\pm$0.37      &6.05$\pm$0.30       &12.96$\pm$0.34\\
\\
Fe K$_\beta$ line Energy (keV)		    &7.077$\pm$0.057	  &7.072$\pm$0.009	  &7.07 (fr)      		 &7.112$\pm$0.038\\
Width of Fe K$_\beta$	(keV)           &0.01$\pm$0.07      &0.01$\pm$0.05      &0.07$\pm$0.07       &0.09$\pm$0.03 \\
Norm.$^c$ of Fe K$_\beta$			        &0.82$\pm$0.10	    &2.89$\pm$0.23	    &1.73$\pm$0.66       &5.79$\pm$1.03 \\
Fe K$_\beta$ eq. width (eV)           &66.94$\pm$13.89		&267.40$\pm$21.60   &46.93$\pm$18.02     &66.08$\pm$11.78 \\
Fe K$_\beta$ line flux$^d$            &0.47$\pm$0.10      &1.18$\pm$0.10      &1.28$\pm$0.49       &3.72$\pm$0.66\\
\\
Flux$^e$ (in 2-10 keV range)          &0.38$\pm$0.09      &0.28$\pm$0.07      &1.61$\pm$0.25       &3.32$\pm$0.30  \\
Flux$^e$ (in 10-70 keV range)         &4.67$\pm$1.44      &4.46$\pm$1.35      &11.49$\pm$1.81      &27.10$\pm$2.22 \\

$\chi^{2}$ (dof)		                  &219 (149)	       	&275 (148)	        &181 (152)		       &262 (142) \\ 
\hline
\end{tabular}
\\
N$_{H}$ = Equivalent hydrogen column density,
$^a$ : photons keV$^{-1}$cm$^{-2}$s$^{-1}$ at 1 keV,
$^b$ : in units of 10$^{39}$ erg s$^{-1}$ {(d/10 kpc)}$^{-2}$ where $d$ is distance to the source,
$^c$ : in units of 10$^{-4}$ photons s$^{-1}$cm$^{-2}$,
$^d$ : in 10$^{-12}$  ergs cm$^{-2}$ s$^{-1}$ unit, 
$^e$ : in 10$^{-10}$  ergs cm$^{-2}$ s$^{-1}$ unit.

\label{spec_par}
\end{table}

\begin{figure}
\centerline{\includegraphics[angle=-90,width=6.5cm]{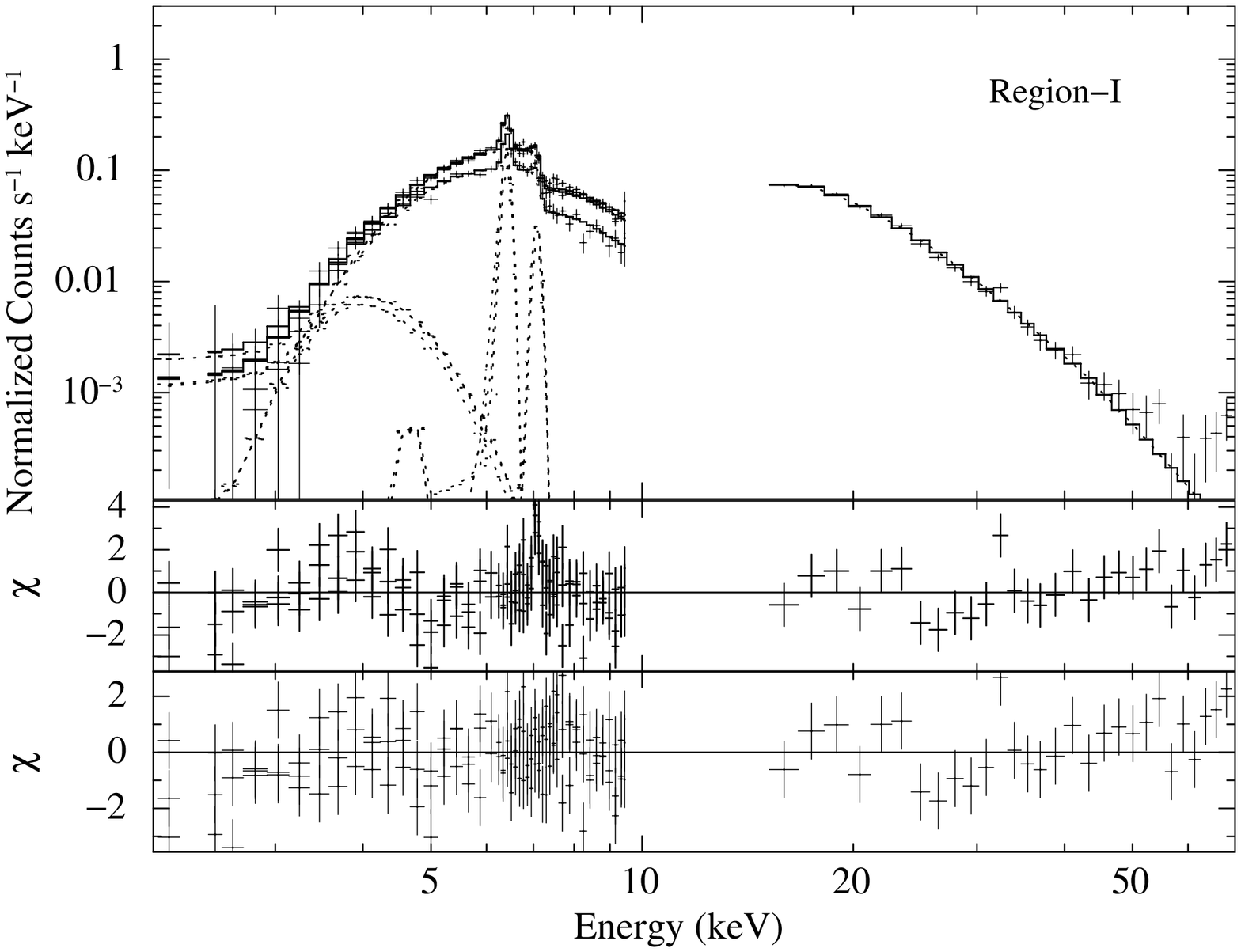}
            \includegraphics[angle=-90,width=6.5cm]{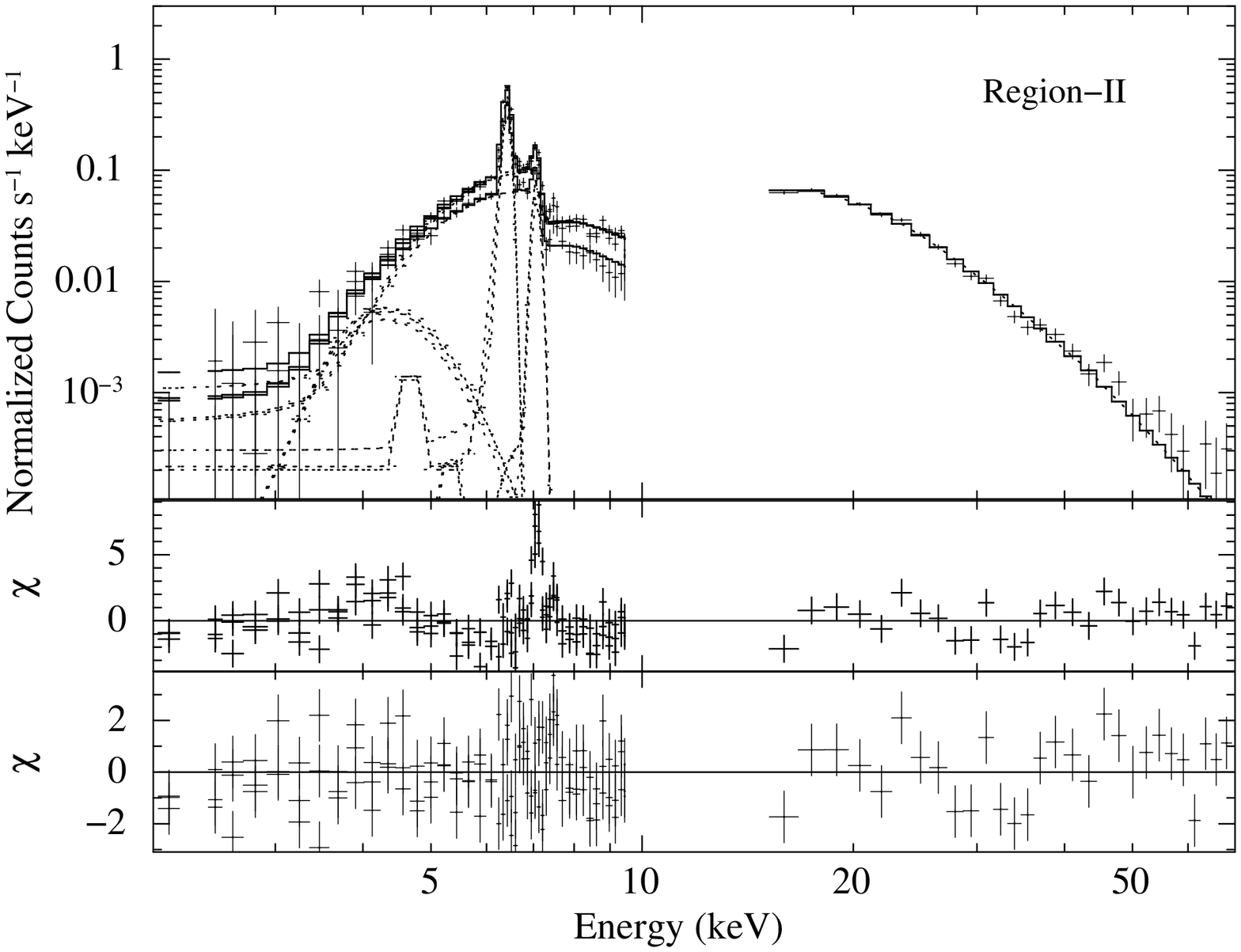}}
\centerline{\includegraphics[angle=-90,width=6.5cm]{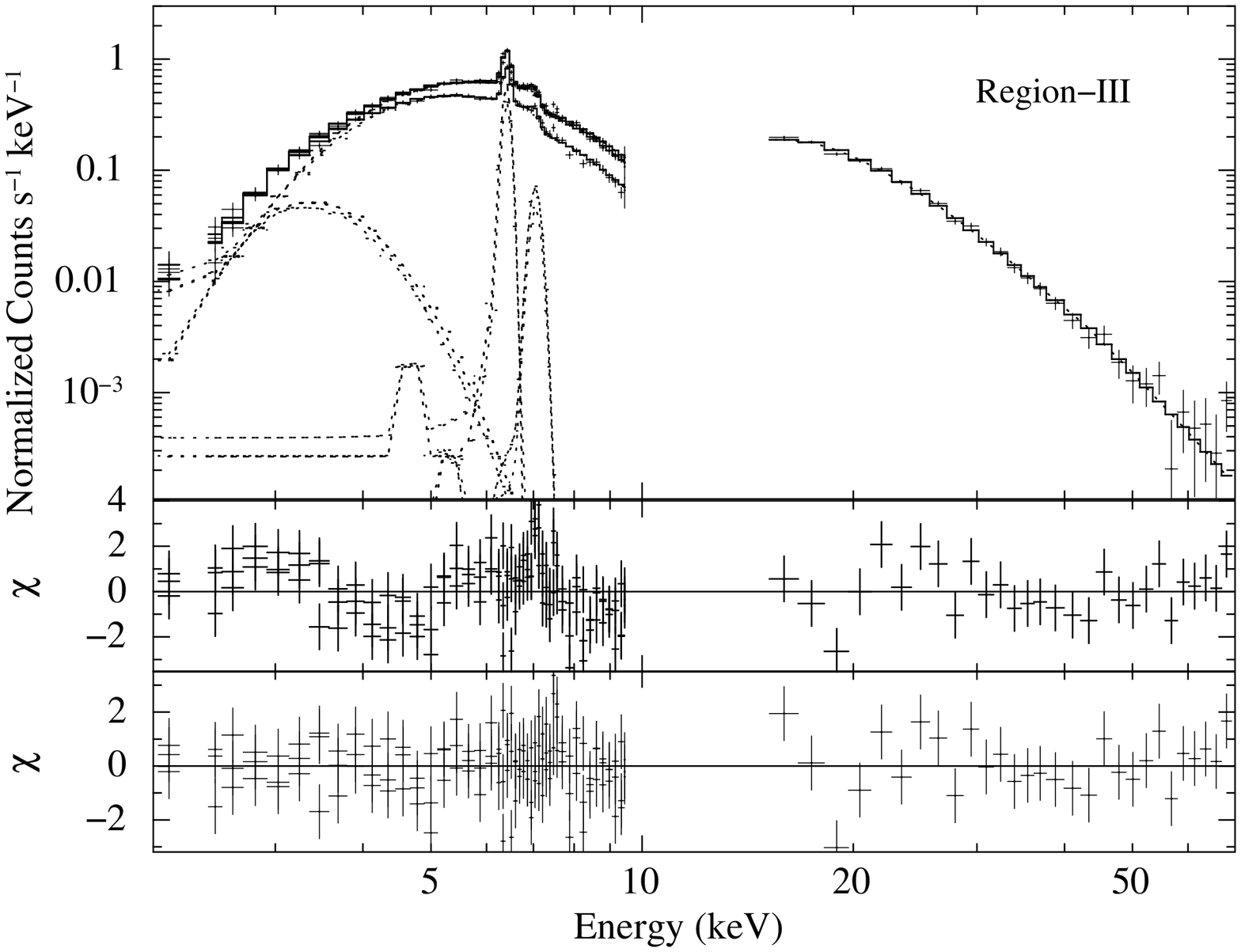}
            \includegraphics[angle=-90,width=6.5cm]{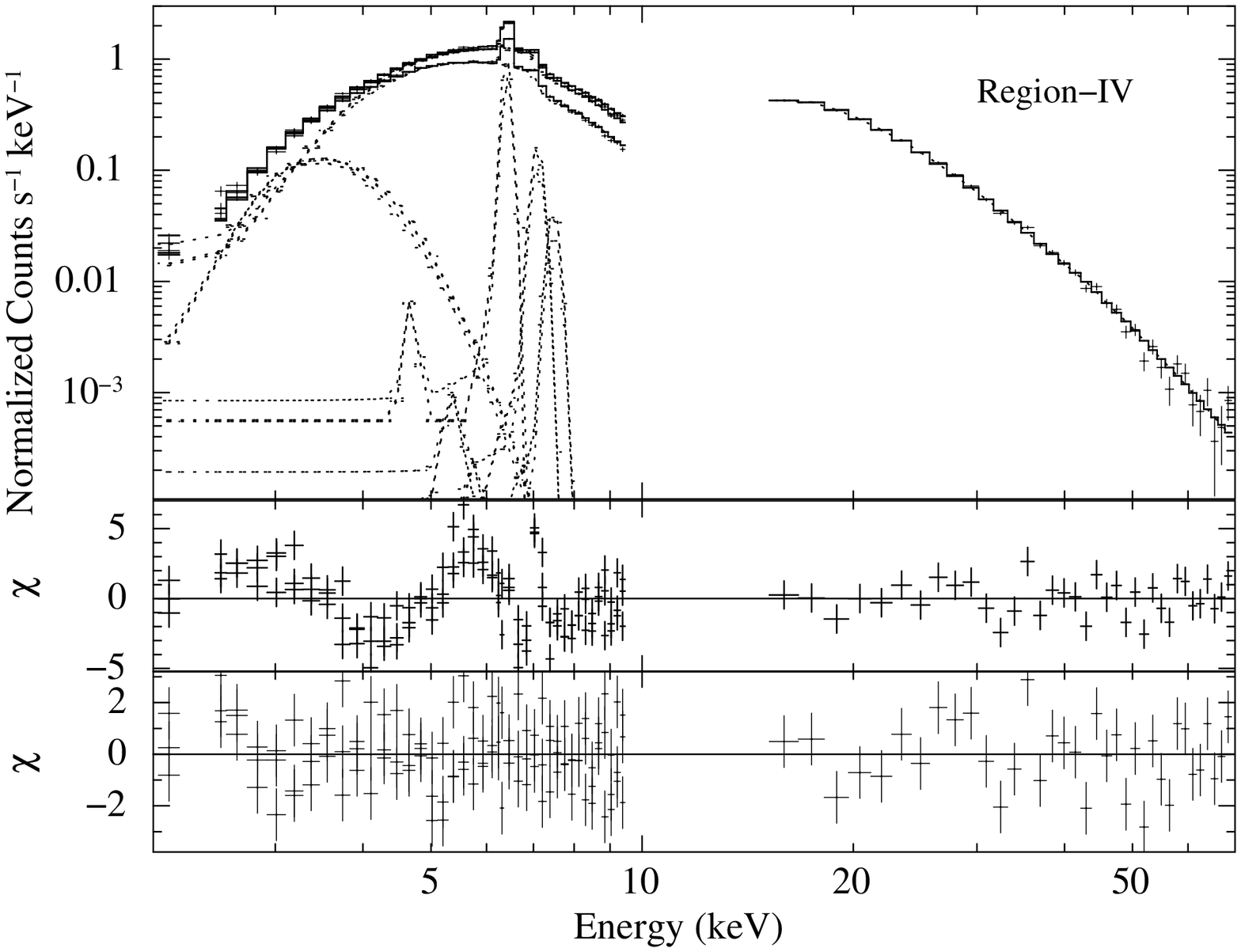}}
\caption{Energy spectrum of OAO~1657-415 at four different flux levels (as shown in
Figure~\ref{xislc}) during the $Suzaku$ observation in 2011 September 26-28. Spectral
data from XIS and HXD/PIN detectors are shown in the top panels of all four regions,
along with the best-fit model comprising of a high energy cutoff power law with a
blackbody component, interstellar absorption and two narrow iron emission lines at
6.4 keV and 7.06 keV. The middle panels show the contribution to the $\chi^2$ for
each energy bin for the high energy cutoff power-law model along with interstellar
absorption and iron K$_\alpha$ emission line. The presence of another emission line
like feature at $\sim$7.06 keV was present in the residuals of each segments.
Along with two iron emission lines, a third line at 7.48 keV which was identified
as Ni K$_\alpha$ line was required to fit the high flux level spectrum of Region-IV.
The contributions of the residuals to the $\chi^2$ for each energy bin for the best-fit
model are shown in the bottom panels for corresponding regions.}
\label{sp_all}
\end{figure}

\begin{figure}
\includegraphics[height=4.9in, width=4.2in, angle=-90]{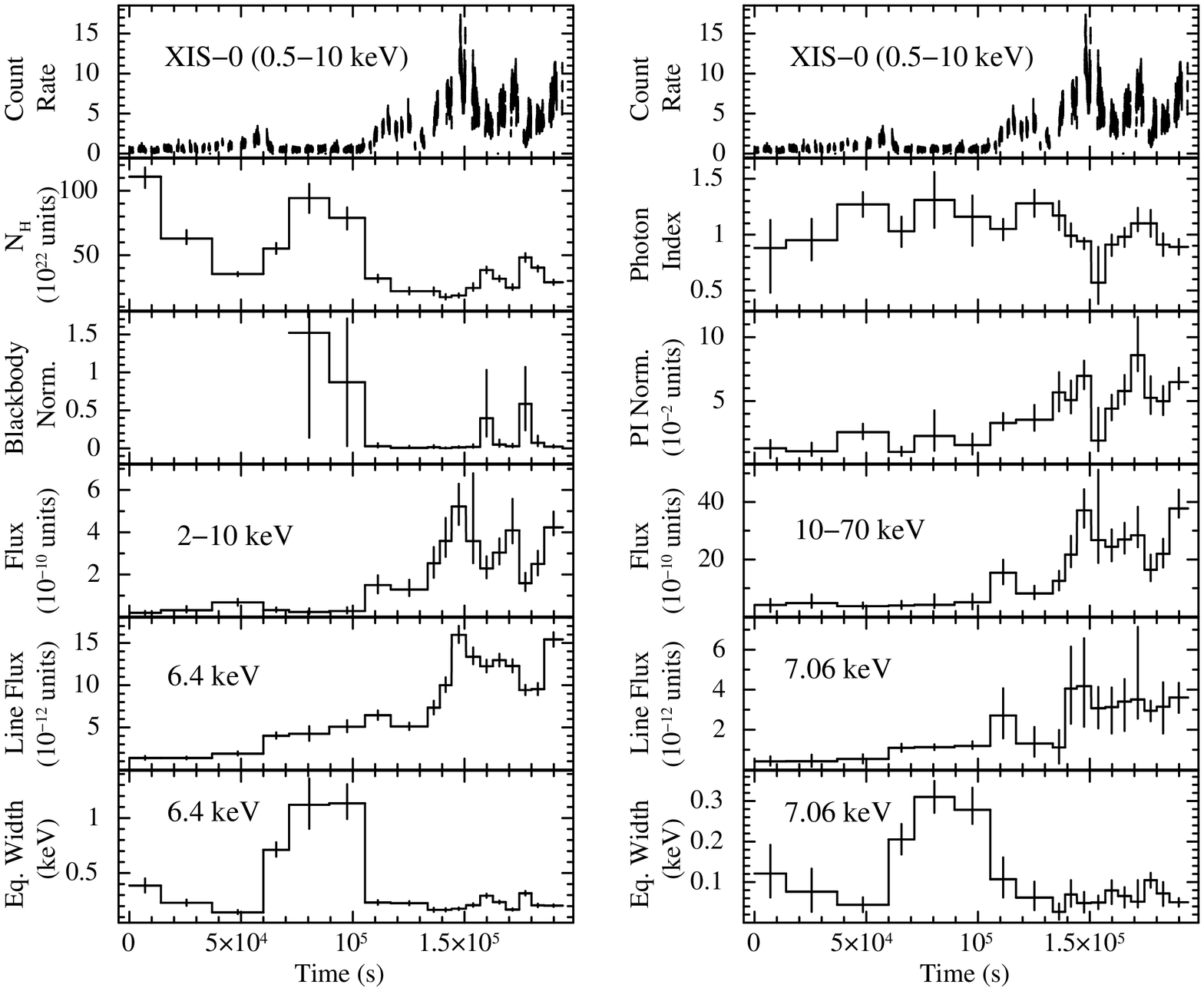}
\caption{Spectral parameters obtained from the time-resolved spectroscopy 
of $Suzaku$ observation of OAO~1657-415. The top panels in both the sides 
show light curves of the pulsar in 0.5-10 keV energy range. The values of 
$N_H$ and power law photon index are shown in second panels in left and 
right side, respectively. The source flux in 2-10 keV (left side) and 
10-70 keV (right side) in 10$^{-10}$ ergs cm$^{-2}$ s$^{-1}$ units are shown 
in third panels. Fourth and fifth panels in both the sides show the line flux 
in 10$^{-12}$ ergs cm$^{-2}$ s$^{-1}$ units and equivalent widths of 6.4 keV 
and 7.06 keV iron emission lines, respectively. The errors shown in the figure 
are estimated for 90$\%$ confidence level.}
\label{sp1}
\end{figure}

The emission line corresponding to the second Gaussian function in the
spectral model was identified to be iron K$_\beta$ fluorescent emission line.
This line was found to be present in the pulsar spectrum during the  
$Suzaku$ observation of the pulsar which has been reported by Pradhan et al. 
(2014). While fitting the spectra of high flux level (Region-IV), we found a 
weak signature of Ni K$_\alpha$ line at 7.48 keV in the residue. Addition of a 
Gaussian component at 7.48 keV improved the $\chi^2$ value from 287 (145 dof) to 
262 (142 dof). The equivalent of Ni K$_\alpha$ line was estimated to be 23.7$\pm$3.5 
eV. This line was also detected in a few other cases such as GX~301-2 (F{\"u}rst et 
al. 2011; Suchy et al. 2012). In the spectral fitting, however, there was no 
signature of presence of earlier reported cyclotron resonance scattering feature 
(CRSF) at $\sim$36 keV (Orlandini et al. 1999). We, therefore, did not include 
CRSF component in the spectral fitting. Simultaneous spectral fitting of XIS 
and HXD/PIN data were carried out for all four regions of the $Suzaku$ observation 
of the pulsar and shown in Figure~\ref{sp_all}. The top panels of the figure 
showed the spectral data along with the best-fit model components for all four 
regions whereas the residuals obtained from the spectral fitting without the 
inclusion of blackbody and 7.06 keV iron emission line in the model are shown 
in middle panels. The bottom panels showed the residuals obtained by adding 
both the components to the spectral model. The best-fit parameters (with 
90$\%$ errors) obtained from simultaneous spectral fitting are given in 
Table~\ref{spec_par}. Among the four regions, the data during Regions-I \& II 
were found to be significantly affected by absorption with the values of absorption
column density of 51$\times$10$^{22}$ atoms cm$^{-2}$ and 76$\times$10$^{22}$ atoms 
cm$^{-2}$, respectively. Though the flux of 6.4 keV and 7.06 keV iron emission lines 
were maximum during the high intensity level of Region-IV, the equivalent width of 
these lines were significantly high during the low flux level of Region-II. 
The continuum flux in 2-70 keV range were found to be comparable during Regions-I 
\& II. For a detailed understanding of the changes in the values of absorption 
column density, equivalent widths and flux of iron emission lines during different
regions, we attempted to do time resolved spectroscopy by dividing the entire 
observation into 18 narrow segments (as marked at the bottom of Figure~\ref{xislc}). 
As described earlier, XIS and HXD/PIN spectra were extracted for each of the 
narrow segments. Previously extracted background spectra and response files were 
used for the spectral fitting of XIS and HXD/PIN data during these narrow segments.
Simultaneous spectral fitting was carried out by using the high energy cutoff power 
law model along with other components as used earlier. Iron line parameters 
were kept free during the spectral fitting with a lower limit of line width at 1 eV. 
Though, the energy and width of both Fe lines were similar during entire observation, 
the energy and width of Fe K$_\beta$ line were found to be high ($\sim$7.28 keV and 
$\sim$0.2 keV, respectively) during segments G $\&$ K. Therefore, the values of line
energy was fixed at 7.07 keV during the segments G and K. The best-fit parameters 
obtained from the simultaneous spectral fitting were plotted in Figure~\ref{sp1} 
along with the source light curve in 0.5-10 keV range in top panels.

\begin{figure}
\centerline{\includegraphics[angle=-90,width=6.5cm]{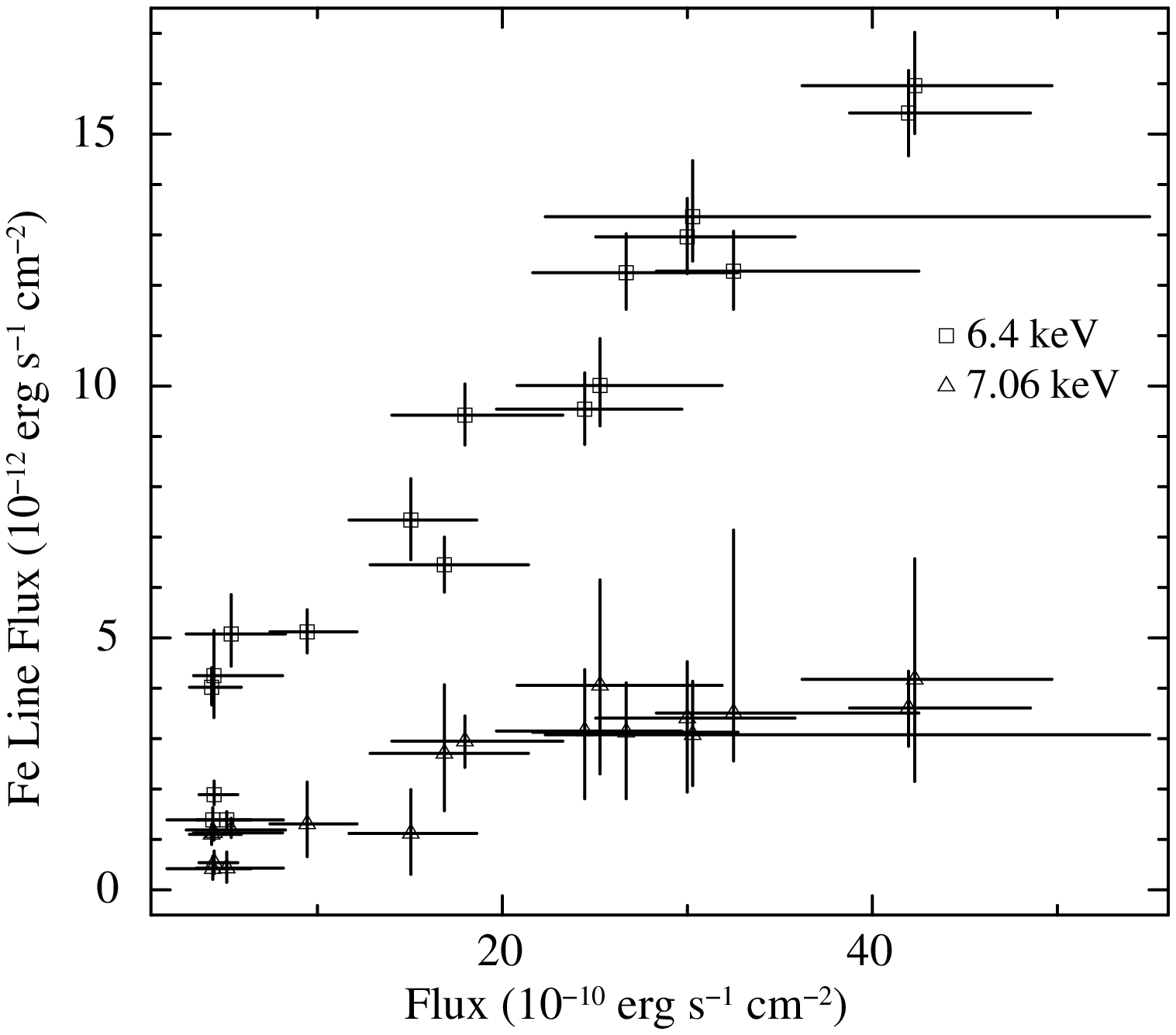}
            \includegraphics[angle=-90,width=6.5cm]{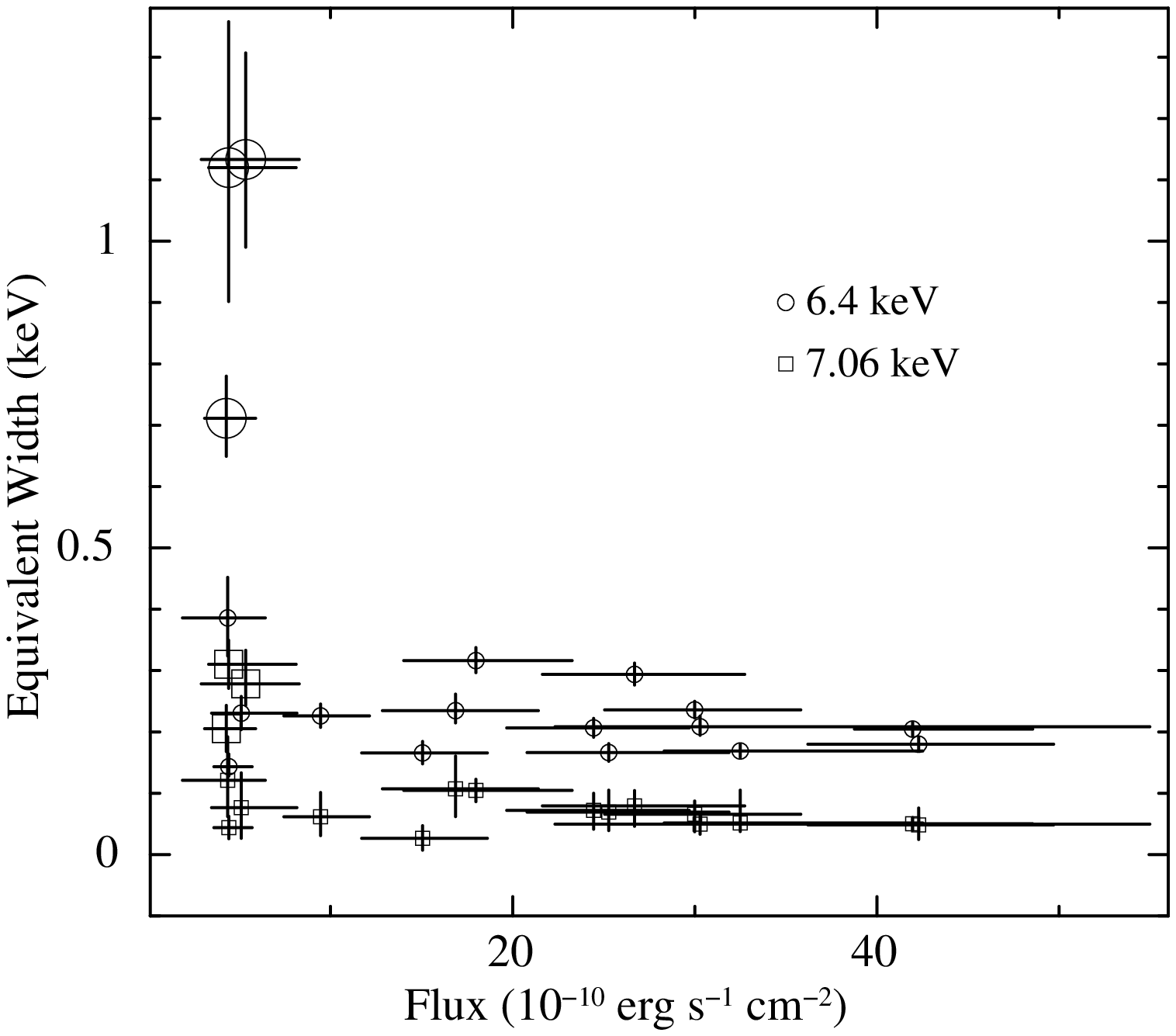}}
\caption{Change in the 6.4 keV and 7.06 keV iron emission line flux (left panel) and
equivalent widths (right panel) with respect to the observed flux in 2-70 keV
energy range. The squares and triangles in left panel represented flux for 6.4 keV and 
7.06 keV iron lines, respectively. In the right panel, the circles and squares showed
the equivalent widths of 6.4 keV and 7.06 keV iron lines, respectively. The markers in
large sizes represented the corresponding values estimated during the low flux segment
of Region-II (in Figure~\ref{xislc}).}
\label{Fe-line}
\end{figure}

The value of equivalent hydrogen column density $N_H$ (second panel; left side of 
Figure~\ref{sp1}) was found to vary in a wide range. It was maximum in the beginning 
of the observation for a short duration (early part of Region-I) which decreased by a 
factor of $\sim$4 and then increased again during the Region-II. Beyond Region-II, the 
value of $N_H$ became low during the high flux segments of Regions-III and IV. The 
change in power law photon index was not significant enough (though a marginal 
decrease was seen towards the end of the observation; second panel - right side) 
to draw any conclusion on spectral state change in the pulsar. The blackbody 
normalization was found to be varying in similar pattern as N$_H$ (Figure~\ref{sp1}). 
However, the normalization of power law component was seen increasing along with the 
source flux. Though, the flux of both the iron emission lines (fifth panels from top) 
increased with the increase in source flux (fourth panels from top), equivalent width 
of these emission lines did not show any such change. The values of equivalent widths 
were significantly high during the low flux level of Region-II. To investigate
the change in iron emission line parameters with the observed continuum flux, we 
plotted flux of iron emission lines and equivalent widths with respect to the 
total flux in 1.8-70 keV energy range (as shown in Figure~\ref{Fe-line}). It was 
found that the flux of both the lines increased continuously with increase in 
the source flux though the increase was significantly high in case of 6.4 keV 
line. However, the equivalent width of both the lines were very high during the 
low flux segment of Region-II (shown in the right panel Figure~\ref{Fe-line} with 
larger size markers) whereas during the rest of the segments, the changes were 
minimal or nearly constant. Unusual high value of equivalent widths for 
both the iron emission lines during Region-II compared to that during Region-I, 
though both were at similar flux level, suggested the presence of significant 
amount of additional matter emitting iron emission lines at that orbital phase 
of the binary pulsar. More observations of OAO~1657-415 with longer exposures 
can provide details of the matter distribution along the orbit of the binary system.

\section{Discussion}
Photons emitted from the magnetic poles of the accretion powered X-ray 
pulsars interact with matter in the surrounding regions such as accretion column,
accretion stream, accretion disk, dust clouds, interstellar medium before 
reaching the observer. These interactions are reflected in the observed spectral 
and timing properties of these objects. Presence of several components such 
as soft X-ray excess, fluorescent emission lines in the spectrum are attributed 
to the evidence of accretion disk formation whereas detection of cyclotron 
absorption features lead to the estimation of surface magnetic field of the 
neutron star. Fluorescent iron emission lines are seen in the spectra of most 
of the X-ray pulsars. X-ray photons emitted from the pulsar interact with the 
ionized/neutral iron atoms emitting characteristic emission lines. These 
interactions occur at various reprocessing sites such as the atmosphere of 
companion star, stellar wind stagnated in shock front, accretion disk. 
Therefore, the investigation of iron emission lines provides powerful 
tool to understand the characteristics of the accretion plasma and their spatial 
distribution in extreme physical conditions around the neutron star. Strength of 
iron emission lines depend on the continuum emission and the density/amount of 
available line emitting materials. The change in observed values of iron emission
line parameters with a short duration over orbital phase implies the presence of 
inhomogeneously distributed matter over the orbit of the binary pulsar.

{\it Suzaku} observation of OAO~1657-415 during out-of-eclipse orbital phase in 
2011 September showed the presence of 6.4 keV and 7.06 keV iron emission lines in 
its spectrum. Although the observed continuum flux of the pulsar was highly variable
during the observation, both the lines were present in the spectrum during entire 
observation. However, strength of these lines were significantly different even during
the extended durations of low intensity levels. During Region-II, the equivalent 
width of the lines were found to be $\sim$10 times higher than that during the
other low flux segment (Region-I). The value of equivalent hydrogen column density 
was also found to be high during Region-II compared to that during Region-I (Table~1). 
Flux of both the lines was found to be directly correlated with the observed continuum 
flux (Figure~\ref{Fe-line}). This indicates the fluorescence origin of iron lines 
from the cool matter surrounding the neutron star (Makishima 1986). The direct 
correlation between the line flux and pulsar continuum flux also suggests that 
the location of reprocessing region is closer to the central source. 

Observed changes in the iron emission line parameters with respect to 
continuum flux led to the interpretation of line emitting regions in the 
accretion powered X-ray pulsars. Investigation of iron emission lines in
the HMXB pulsar Cen~X-3 during eclipse-ingress, eclipse, eclipse-egress
and out-of-eclipse phases confirmed that the Fe K$_\alpha$ line is being 
originated from regions close to neutron star whereas 6.7 keV and 6.97 keV 
lines are produced in a region that is far from the neutron star, probably 
in the highly photo-ionized wind of the companion star or in the accretion 
disk corona (Ebisawa et al. 1996; Naik et al. 2011; Naik \& Paul 2012). 
Detection of pulsation in 6.4 keV iron emission line in some cases such as 
Vela~X-1 (Choi et al. 1996), Her~X-1 (Zane et al. 2004) etc. confirmed the 
reprocessing region to be close to the neutron star. ASCA observations of 
OAO~1657-415 confirmed that the origin of the 6.4 keV iron fluorescent emission 
line is originated from regions closer to the neutron star whereas the 6.9 keV 
recombination line is originated from the extended ionized stellar wind of the 
optical companion star (Audely et al. 2006). Dependence of iron emission line 
parameters with source continuum flux had also been investigated in a few X-ray 
pulsars such as LMC~X-4 and Her~X-1 (Naik \& Paul 2003). A direct correlation 
between the flux of iron emission line and continuum flux was detected in both 
the pulsars whereas no systematic variation was detected in the values of 
equivalent width and source flux. In case of both the pulsars, the flux modulation 
at super-orbital periods were interpreted as due to the presence of precessing 
tilted accretion disk that blocks the direct X-ray beam from the pulsar. Considering 
the similarities in the changes in the iron line parameters with source flux in 
Cen~X-3, LMC~X-4, Her~X-1, we suggest that the flux variation at different orbital 
phases of OAO~1657-415, as seen in the 2011 $Suzaku$ observation of the pulsar, 
can be explained by the presence of additional dense absorbing matter at corresponding 
orbital phases. Further observations of OAO~1657-415 would confirm whether the segment
of low flux level (due to obscuration/absorption of direct X-rays from the pulsar)
repeatedly present (possibly because of the presence of structures in the close
proximity of the pulsar) or a temporary one due to the presence of clump of matter. 

The pulsar OAO~1657-415 showed significant flux variability during the 
$Suzaku$ observation in 2011 September. Though the observation was carried out 
during out-of-eclipse phase of 10.44 d orbital period, there were two elongated 
segments of low flux levels followed by regions of high flux level. Similar flux 
variability was seen in the light curve of another HMXB pulsar Cen~X-3 (Naik et 
al. 2011) obtained from {\it Suzaku} observation in 2008 December covering nearly 
one orbital period. Segments of low and high flux levels occurred more frequently 
during the {\it Suzaku} observation of the pulsar. The observed changes in the 
source flux as well as the value of absorption column density during the $Suzaku$
observation of Cen~X-3 is found to be comparable to that observed during the $Suzaku$
observation of OAO~1657-415. 

Using the same observation used in the present work, Pradhan et al. (2014) have 
interpreted the flux variability of OAO~1657-415 as due to the presence of dense clump 
of matter at certain orbital phases. In the spectral fitting, Pradhan et al. (2014) used
data from all three XIS and PIN covering in 3-70 keV range. In the present study, however, 
we have included the low energy data in our spectral fitting for the better estimation of
the value of equivalent hydrogen column density at different orbital phases of the binary
pulsar. In our fitting, the pulsar spectrum in 1.8-70 keV was well described with high 
energy cutoff power-law model along with other components for interstellar absorption and
iron emission lines. Though, Pradhan et al. (2014) excluded data below 3 keV, partial 
absorbing component was required in their spectral fitting for two segments. However, 
in our spectral fitting in 1.8-70 keV range, the additional absorption component was
not required during any of the narrow data segments shown in Figure~2. The interesting
thing was the pattern of variation in the value of absorption column density was similar 
in both the cases (Figure~4 of our present work and Figure~8 of the work of Pradhan et al.
2014). Apart from the value of N$_H$, the equivalent width of both iron emission lines
were found to be maximum during the second extended low segment. Though Pradhan et al.
(2014) needed a cyclotron resonance scattering feature (CRSF) component in their spectral
fitting to the data during both the extended low segments, no such component was required
while fitting the data in the present work. Using the same {\it Suzaku} observation of the
pulsar, here we are investigating in detail the characteristic physical parameters of the 
clumps causing the flares and low flux segments in the light curve.

Significant flux variability in OAO~1657-415 during $Suzaku$ observation can
be explained by using clumpy stellar wind model (Feldmeier et al. 2003; Oskinova 
et al. 2007). The accretion of clumps of matter onto the pulsar causes increase 
in source luminosity by producing flares/flare-like episodes in X-ray emission. 
The obstruction/absorption of X-ray photons by dense clumps of matter in the 
line-of-sight can also cause segments of low flux levels in X-ray light curve.
Considering inhomogeneous distribution of matter in the stellar wind, the 
luminosity of such wind-powered HMXB pulsars depends on the density and velocity 
of the stellar wind through the relation (Bondi \& Hoyle 1944)
\begin{equation}
L_x \propto \rho v^{-3} 
\end{equation}
where $\rho$ is density and $v$ is the wind velocity as described in $\beta$-velocity 
law (Castor et al. 1975). Any fluctuation in either density or 
velocity profile produces the time variability in the source luminosity. 

The clumpy wind model can explain the observed variability in circumstellar 
absorption and source flux. The physical parameters of the clumpy wind can be 
estimated from the observed time variability in the light curve of 
OAO~1657-415 using this model. Figure-$\ref{xislc}$ shows the presence of a 
series of flares of duration of $\gtrsim$10~ks. Considering a flare of duration 
t$_f\sim10$~ks and relative wind velocity v$_{rel}\sim250$~km~s$^{-1}$ 
(taking an approximation of terminal velocity similar to the relative velocity; 
Mason et al. 2012), the radius of the clump can be 
\begin{equation}
R_c\simeq \frac {v_{rel} \ t_f}  {2}
=1.25\times10^{11} \left ( \frac {v_{rel}}{250 \ \mathrm {km \ s^{-1}}} \right ) \left ( \frac {t_f}{10 \ \mathrm {ks}} \right )  \mathrm{cm}
\end{equation}

Gravitational influence of the neutron star on the clumps becomes effective at the 
accretion radius r$_{acc}\sim$2GM$_{ns}$/v$_{rel}^2$. As a clump of matter crosses this 
radius, it gets accreted toward the neutron star. The accretion radius for OAO~1657-415
is calculated to be $\sim$6$\times$10$^{11}$~cm by assuming typical parameters of the
neutron star and relative stellar wind velocity of 250~km~s$^{-1}$. Accretion radius also 
imposes a constrain on upper limit of the radius of a clump that can be accreted entirely 
onto the neutron star. Mass of the accreted clump can be estimated by comparing the 
gravitational potential energy of clump to the energy released due to its accretion 
onto the neutron star (assuming efficiency $\eta$ as 0.1) through the relation 
(Zel'dovich \& Shakura 1969)
\begin{equation}
M_c = \frac{L_x t_f R_{ns}}{\eta  GM_{ns}} = 5\times 10^{21} \left ( \frac {L_x}{10^{37} \ \mathrm {erg \ s^{-1}}} \right ) \left ( \frac {t_f}{10 \ \mathrm {ks}} \right )  ~ \mathrm {g}
\end{equation}
where M$_{ns}$, R$_{ns}$ and L$_x$ are mass, radius and X-ray luminosity of the neutron star,
respectively. Therefore, for a given set of physical parameters of the spherical clump, its 
mean density (n$_c$) and radial column density (N$_c$) can be estimated as
\begin{multline}\label{md}
n_c  = \frac {3M_c}{4 \pi m_p R_c^3} = 4\times 10^{11}  \left ( \frac {L_x}{10^{37} \ \mathrm {erg \ s^{-1}}} \right ) 
\times \left ( \frac {t_f}{10 \ \mathrm {ks}} \right )^{-2} \left ( \frac {v_{rel}}{250 \ \mathrm {km \ s^{-1}}} \right )^{-3}  \mathrm {cm^{-3}}
\end{multline}

\begin{multline}\label{rcd}
N_c=n_c R_c= 5\times 10^{22} \left ( \frac {L_x}{10^{37} \ \mathrm {erg \ s^{-1}}} \right ) \times \left ( \frac {t_f}{10 \ \mathrm {ks}} \right )^{-1} \left ( \frac {v_{rel}}{250 \ \mathrm {km \ s^{-1}}} \right )^{-2}  \mathrm {cm^{-2}}
\end{multline} 

Using the typical parameters in equation ($\ref{md}$), the mean density of a clump can 
be estimated to be 4$\times$10$^{11}$~cm$^{-3}$. However, the mean number density of the 
stellar wind around the neutron star is calculated to be n$\sim$4$\times$10$^{10}$~cm$^{-3}$  
by using the $\beta$-velocity law (Castor et al. 1975) and the wind parameters 
from Mason et al. (2012). Stellar wind density is found to be  an order of magnitude lower 
than the value obtained from equation~($\ref{md}$). 

In the present {\it Suzaku} observation, the duration and luminosity of the flare 
(segments G \& H) are about $\sim$25~ks and 0.66$\times$10$^{37}$~erg~s$^{-1}$, respectively.
Using these values in Equation~\ref{rcd}, the clump column density ($N_c$) is estimated to be 
1.3$\times$10$^{22}$~cm$^{-2}$. However, the difference in the observed value of N$_H$ (from
spectral fitting) for segments F (prior to the flare) and G \& H (during flare) is found to be 
52$\times$10$^{22}$~cm$^{-2}$. This value corresponds to the column density of clump that produced 
a flare of $\sim$25~ks duration (segments G \& H). However, it is found to be $\sim$40 times more 
than the expected column density ($N_c$). In case of another flare (segments I to L) with 
duration of $\sim$24~ks and luminosity of 1.38$\times$10$^{37}$~erg~s$^{-1}$, the clump column 
density ($N_c$) is estimated to be 2.9$\times$10$^{22}$~cm$^{-2}$. From spectral fitting, the
difference in $N_H$ between the flare (segments from I to L) and the segment prior to the flare 
(segment H) is calculated to be 1.4$\times$10$^{22}$~cm$^{-2}$ which is comparable to the estimated
value $N_c$. For the flare starting from segment N to segment P, the duration and luminosity are 
$\sim$12~ks and 1.31$\times$10$^{37}$~erg~s$^{-1}$, respectively. Using these values, the clump 
column density ($N_c$) is estimated to be 5.4$\times$10$^{22}$~cm$^{-2}$. However, as calculated
earlier, the clump density from spectral fitting is found to be  3.6$\times$10$^{22}$~cm$^{-2}$
which is also close to $N_c$. From these examples, it is found that the values of observed and 
calculated clump column densities are comparable. This supports the suggestion that the observed 
flux variability in the pulsar light curve can be explained by using clumpy wind hypothesis. 

However, in case of the flare during segments G \& H, the estimated clump column density 
from spectral fitting is quite high compared to the expected column density calculated by 
using Equation~\ref{rcd} in case of a spherically symmetric clump of homogeneous density.
The enhancement in the value of column density compared to the standard wind density can 
occur because of the instability in variable and highly structured stellar wind. 
As the supersonic stellar wind interacts with the neutron star, a bow shock of compressed 
gas forms near r$_{acc}$ that may cause enhancement in the column density. Numerical simulation 
results for the wind-fed sources indicate that the regions inside the non-steady accretion wake 
consist of  dense filaments in which the density reaches up to $\sim$100 times more as compared 
to undisturbed stellar wind (Blondin et al. 1990). Such dense clump or filament can absorb the 
X-ray photons up to higher energies and results the low flux levels in the light curve. 

Interaction of X-ray photons either in dense filaments or stagnated shock front can contribute 
to higher value of the equivalent width. Inoue (1985) performed Monte Carlo simulation for the 
fluorescence emission from the neutral atom as a function of absorber density. Different model 
components were considered in the above study. A model with a combination of direct and scattered 
continuum component explained the possibility of high equivalent width (up to order of keV) with 
variation in N$_H$. If the absorbing matter situated directly between the observer and the X-ray 
source, then the equivalent width for emission lines was found to be monotonically proportional 
to the column density. As a result the high equivalent width (more than 100 eV) for 6.4 keV line 
was seen at N$_H$ $>$ 10$^{23}$~cm$^{-2}$.   
 
In summary, we interpret our results based on the clumpy wind hypothesis. The extended low flux 
levels of Region-I \& II can be interpreted as the presence of dense clump or filament of 
the compressed gas. The high equivalent width of Fe K$_\alpha$ and Fe K$_\beta$ lines can be 
understood by the result obtain from Monte Carlo simulation (Inoue 1985) in which the dense 
clump of matter obscures the direct X-rays and contributes to the high value of equivalent 
width as observed in Region-II. During the orbital phase 0.19-0.24 (Region-II), the clump 
could be situated along the line-of-sight. As the clump passes Region-II the absorption column 
density decreases and causes decrease in the equivalent width of iron emission lines. Lifetime 
of this clump can be similar to duration of Region-II. Furthermore, decrease in the absorption 
column density and increase in the source flux in Region-III \& Region-IV suggest the 
obstruction or either the accretion of matter by relatively low dense or off line clumps. 

\subsection{Location of the Fe fluorescence line-emitting region}
To investigate the ionization state of matter (iron) in line emitting region, 
the ratio of Fe K$_\beta$ line flux to that of Fe K$_\alpha$ line was calculated 
for the entire $Suzaku$ observation of OAO~1657-415. It was found that the value 
of flux ratio varies between 0.15$\pm$0.11 and 0.42$\pm$0.18 for all segments. 
The value of line flux ratio for each segment was found to be higher than the 
theoretically predicted value ($\sim$0.13) for neutral iron atom in optical thin 
medium (Kaastra \& Mewe 1993; Palmeri et al. 2003; Han \& Demir 2009). This suggests 
that the line emitting region can be a mixture of neutral and ionized iron atoms. 
The observed line energy and flux ratio indicate the possible existence of ionized 
iron atoms in ionization state between Fe VIII to Fe XVIII 
(Palmeri et al.  2003; Mendoza et al. 2004 and Kallman et al.  2004). 
Ionization of the iron atoms is characterized by the ionization parameter 
$\xi$=L/nr$^2$ and its value should be $\xi\sim$10$^{2.5}$~erg cm s$^{-1}$ 
for ionization state below Fe XVIII (Tarter et al. 1969; Ebisawa et al. 1996). 
However, the iron atoms are expected to be fully ionized at 
$\xi\ge$10$^{3}$~erg cm s$^{-1}$ (Kallman \& McCray 1982). The peak population 
of ion at Fe XVIII can expected to be $\sim$0.25 (Ebisawa et al. 1996).
For a luminosity of 10$^{37}$~erg~s$^{-1}$ and column density of 
10$^{23}$$\le$N$_H$$\le$ 10$^{24}$~cm$^{-2}$, the location of ionization region 
is found to be in range of (0.3-3)$\times$10$^{11}$~cm from the neutron star. 
Using this method, we found that the location of iron ionization region lies 
within or closer to the accretion radius r$_{acc}$. Our result agrees with 
that obtained from earlier studies in which the location of iron fluorescence 
region was estimated to be within the region of 5.7$\times$10$^{11}$~cm 
(19 lt-sec) (Audley et al. 2006).  

\section{Conclusions}
We have analyzed the timing and spectral properties of OAO~1657-415 
during a long Suzaku observation in 2011 September. The pulsar was 
found to be exhibiting low and high flux levels during the observation. 
The segments of low flux levels are seen up to as high as $\sim$50 keV 
in the light curves of the pulsar. The spectrum of the pulsar in 1.8-70 keV 
range was found to be well described by the high energy cutoff model. Iron 
fluorescent emission lines at 6.4 keV and 7.06 keV were detected in the 
spectrum of the pulsar through out the low and high flux levels. Flux of 
the iron emission lines was found to be variable with the continuum flux 
whereas the equivalent width was nearly constant with flux except for 
Region-II. The correlation between observed line flux and continuum flux 
suggests the fluorescent origin of the iron lines from the regions 
near to the neutron star. It is evident from the line energy 
and flux ratios of Fe K$_\beta$ and Fe K$_\alpha$ lines that line emitting 
regions consist of a mixture of neutral and ionized iron atoms. We estimated 
distance of the fluorescence region and found that region lies within the 
accretion radius. Considering the high value of absorption density during 
the segments of low flux level and high values of iron line equivalent 
widths, we can explain that the low flux segments in OAO~1657-415 are 
due to the presence of dense clumpy matter at certain orbital phases of 
the binary system.

\section*{Acknowledgments}
We thank the referee for her/his constructive comments and suggestions 
that improved the content of the paper. The research work at Physical 
Research Laboratory is funded by the Department of Space, Government of India. 
The authors would like to thank all the members of {\it Suzaku} for their contributions 
in the instrument preparation, spacecraft operation, software development, and in-orbit 
instrumental calibration. This research has made use of data obtained through HEASARC 
Online Service, provided by NASA/GSFC, in support of NASA High Energy Astrophysics Programs.


\begin{thebibliography}{}
\bibitem[]{}Audley M. D., Nagase F., Mitsuda K., Angelini L., Kelley R. L., 2006, MNRAS, 367, 1147
\bibitem[]{}Barnstedt J., Staubert R., Santangelo A., et al., 2008, A\&A, 486, 293
\bibitem[]{}Baykal A., 1997, A\&A, 319, 515
\bibitem[]{}Baykal A., 2000, MNRAS, 313, 637
\bibitem[]{}Bildsten L., Chakrabarty D., Chiu J., et al., 1997, ApJS, 113, 367
\bibitem[]{}Blondin J. M., Kallman T. R., Fryxell B. A., Taam R. E., 1990, ApJ, 356, 591
\bibitem[]{}Bondi H., Hoyle F., 1944, MNRAS, 104, 273
\bibitem[]{}Castor J. I., Abbott D. C., Klein R. I., 1975, ApJ, 195, 157
\bibitem[]{}Chakrabarty D., Grunsfeld J. M., Prince T. A., et al., 1993, ApJ, 403, L33
\bibitem[]{}Chakrabarty D., Wang Z., Juett A. M., Lee J. C., Roche P., 2002, ApJ, 573, 789
\bibitem[]{}Choi C. S., Dotani T., Day C. S. R.,  Nagase F., 1996, ApJ, 471, 447
\bibitem[]{}Corbet R. H. D., 1986, MNRAS, 220, 1047
\bibitem[]{}Ebisawa K., Day C. S. R., Kallman T. R., et al., 1996, PASJ, 48, 425
\bibitem[]{}Feldmeier A., Oskinova L., Hamann W., 2003, A\&A, 403, 217
\bibitem[]{}{F{\"u}rst} F., Suchy S., Kreykenbohm I., et al., 2011, A\&A, 535, 9 
\bibitem[]{}Han I., Demir L., 2009, Phys. Rev. A, 80, 052503
\bibitem[]{}Inoue H., 1985, Space Sci. Rev., 40, 317
\bibitem[]{}Jenke P. A., Finger M. H., Wilson-Hodge C. A., Camero-Arranz A., 2012, ApJ, 759, 124
\bibitem[]{}Kaastra J. S., Mewe R., 1993, A\&AS, 97, 443
\bibitem[]{}Kallman T. R., McCray R. 1982, ApJ, 50, 263 
\bibitem[]{}Kallman T. R., Palmeri P., Bautista M. A., et al., 2004, ApJS, 155, 675
\bibitem[]{}Kamata Y., Koyama K., Tawara Y., et al., 1990, PASJ, 42, 785
\bibitem[]{}Kokubun M., Makishima K., Takahashi T., et al., 2007, PASJ, 59, 53
\bibitem[]{}Koyama K., Tsunemi H., Dotani T., et al., 2007, PASJ, 59, 23
\bibitem[]{}Makishima K., 1986, The Physics of Accretion onto Compact Objects, 
                         ed. K. P. Mason, M. G. Watson, N. E. White, Lecture Notes in Physics, 
                         Berlin: Springer, 266, 249 
\bibitem[]{}Mason A. B., Clark J. S., Norton A. J., Negueruela I., Roche P., 2009, A\&A, 505, 281 
\bibitem[]{}Mason A. B., Clark J. S., Norton A. J., et al., 2012, MNRAS, 422, 199 
\bibitem[]{}Mendoza C., Kallman T. R., Bautista M. A., Palmeri P., 2004, A\&A, 414, 377
\bibitem[]{}Mitsuda K., et al., 2007, PASJ, 59, 1
\bibitem[]{}Naik S., Paul B., 2003, A\&A, 401, 265
\bibitem[]{}Naik S., Paul B. 2012, BASI, 40, 503
\bibitem[]{}Naik S., Paul B., Ali Z., 2011, ApJ, 737, 79
\bibitem[]{}Orlandini M., dal Fiume D., del Sordo S., et al., 1999, A\&A, 349, L9 
\bibitem[]{}Oskinova L. M., Hamann W., Feldmeier A., 2007, A\&A, 476, 1331
\bibitem[]{}Palmeri P., Mendoza C., Kallman T. R., et al., 2003, A\&A, 410, 359
\bibitem[]{}Parmar A. N., Branduardi-Raymont G., Pollard G. S. G., et al., 1980, MNRAS, 193, 49
\bibitem[]{}Pradhan P., Maitra C., Paul B., Islam N., Paul B. C., 2014, MNRAS, 442, 2691	
\bibitem[]{}Polidan R. S., Pollard G. S. G., Sanford P. W., Locke M. C., 1978, Nature, 275, 296
\bibitem[]{}Suchy S., {F{\"u}rst} F.,  Pottschmidt K., et al. 2012, ApJ, 745, 124 
\bibitem[]{}Takahashi T., Keiichi A., Endo M., et al., 2007, PASJ, 59, 35
\bibitem[]{}Tarter C. B.  Salpeter E. E., 1969, ApJ, 156, 953
\bibitem[]{}White N. E.,  Pravdo S. H., 1979, ApJ, 233, L121
\bibitem[]{}Zane S., Ramsay G., Jimenes-Garate M. A., et al., 2004, MNRAS, 350, 506
\bibitem[]{}Zel'dovich Ya. B., Shakura N. I., 1969, SvA, 13, 175

\end{thebibliography}
\end{document}